\documentclass[aps,fleqn,preprint]{revtex4-1}

\usepackage{amsmath}
\usepackage{graphicx}
\usepackage{epsfig}
\usepackage{subfigure}
\usepackage{amsfonts}
\usepackage{amssymb}
\usepackage{amsthm}
\usepackage{newlfont}
\usepackage{epstopdf}

\usepackage[utf8]{inputenc}

\usepackage{dcolumn}
\usepackage{bm}

\usepackage[section] {placeins}

\begin{document}

\title{Numerical solution to the time-independent Gross-Pitaevskii equation }

\author{Tsogbayar Tsednee}
\author{Banzragch Tsednee}
\author{Tsookhuu Khinayat}
\affiliation{Institute of Physics and Technology, Mongolian Academy of Sciences, Peace Ave 54B, 13330 Ulaanbaatar, Mongolia }

\begin{abstract}

We solve the time-independent Gross-Pitaevskii equation modeling the Bose-Einstein condensate trapped in an anistropic harmonic potential using a pseudospectral method. 
Numerically obtained values for an energy and a chemical potential for the condensate with positive and negative scattering length have been compared with those from the literature. The results show that they are in good agreement when an atomic interaction is not too strong.

\end{abstract}

\pacs{Valid PACS appear here}
                             
\keywords{Bose-Einstein condensate, nonlinear Schr\"{o}dinger equation, chemical potential, rubidium, lithium,}                             

\maketitle

\section{Introduction}

When the thermal de Broglie wavelength exceeds the mean spacing between identical boson-particles, bosons are stimulated by presence of other bosons in the lowest energy state to occupy that state  as well, resulting in macroscopic occupation of a single quantum state \cite{Bose24, Einstein25}. This phenomenon is named the Bose-Einstein condensation and the condensate that forms constitutes a macroscopic quantum mechanical object. This theoretical prediction had been confirmed experimentally 70 years later, particularly  for $^{87}\mbox{Rb}$~\cite{Anderson95}, $^{7}\mbox{Li}$~\cite{Bradley95} and $^{23}\mbox{Na}$~\cite{Davis95}. The vapors of alkali atoms employed in the experiments are very dilute, so one can expect that the two-body collision accounting for by the knowledge of the $s$-wave scattering length might be dominate. This also implies that the Gross-Pitaevskii theory~\cite{Pitaevskii61, Gross61} for weakly interacting bosons can be suitable for the system which can be simulated to be confined in an isotropic~\cite{Edwards95,Ruprecht95} and an anisotropic~\cite{Dalfovo96,Baym96} traps. 

In this work we solve the time-independent Gross-Pitaevskii equation (GPE) for $N$ alkali atoms in an anisotropic trap. We compute the condensate wave function at $T=0$ for bosons interacting through positive and negative scattering lengths and obtain the chemical potential and energy as a function of $N$. Numerical method we choose to solve the GPE is a pseudospectral method, which we had applied successfully in the past~\cite{Tsogbayar13}.  

The paper is organized as follows. In Section 2, we show the formalism of the Gross-Pitaevskii theory for the anisotropic trap. In Section 3 we give a brief discussion of a pseudospectral approach for the 3D problem. In Section 4 we present the numerical results for the two cases of positive $^{87}\mbox{Rb}$ and negative $^{7}\mbox{Li}$ scattering lengths. Then a conclusion follows.  

\section{Gross-Pitaevskii theory for trapped bosons}

The mean field theory for a dilute assembly of bosons at $T=0$ results in an effective nonlinear Schr\"{o}dinger equation
for the condensate's wave function. This equation, the Gross-Pitaevskii, nonlinear Schr\"{o}dinger equation for condensed bosons has a form:
\begin{eqnarray}\label{GPE}
i\hbar \frac{\partial \Psi (\mathbf{r}, t)}{\partial t} = \Big( -\frac{\hbar^{2}}{2 m}\nabla^{2} + V_{ext} + \frac{4\pi \hbar^{2} a N}{m} |\Psi(\mathbf{r}, t)|^{2} \Big) \Psi(\mathbf{r}, t).   
\end{eqnarray}
Here $\Psi(\mathbf{r}, t)$ is the Bose-Einstein condensate (BEC) wave function, (also called the order parameter), $m$ is the mass of boson, $V_{ext}$ is an external confining potential (trap), $a$ is the $s$-wave scattering length and $N$ is the number of bosons in the condensate.   
  
A stationary solution $\Psi(\mathbf{r}, t) = e^{i\mu t/\hbar} \psi(\mathbf{r})$ obeys 
\begin{eqnarray}\label{TIGPE}
 \Big[-\frac{\hbar^{2}}{2 m} \nabla^{2} + V_{ext}(\mathbf{r}) + V_{mf}(\mathbf{r}) \Big]\psi(\mathbf{r}) = \mu \psi(\mathbf{r}),
\end{eqnarray}
where the mean-field (mf) potential is $V_{mf} =  \frac{4\pi \hbar^{2} a N}{m}  |\psi(\mathbf{r}) |^{2}$.
Once this equation is solved the chemical potential $\mu$ is known and the free energy can be calculated using 
\begin{eqnarray}\label{free_en}
  E = \mu - \frac{1}{2} \langle V_{mf} \rangle = \int \psi(\mathbf{r})^{\ast} \Big( -\frac{\hbar^{2}}{2 m} \nabla^{2} + V_{ext}(\mathbf{r}) + \langle V_{mf}\rangle  \Big) \psi(\mathbf{r}) d\mathbf{r}. 
\end{eqnarray}  
Since $\psi$ and $V_{mf}$ in equation (\ref{TIGPE}) depend on each other, the GP equation must be solved self-consistently. One first uses an initial guess for the wave function $\psi$ to calculate $V_{mf}$ using equation (\ref{mfp}). This value is then employed in equation (\ref{TIGPE}) to obtain a new $\psi$, which is then used to calculate $V_{mf}$ again. This process is repeated until self-consistency is reached.   

In our calculation we use a following anisotropic harmonic oscillator potential:
\begin{eqnarray}\label{mfp}
 V_{ext}(x,y,z) = \frac{m}{2} \omega^{2}_{x} x^{2} +  \frac{m}{2} \omega^{2}_{y} y^{2} +  \frac{m}{2} \omega^{2}_{z} z^{2}.
\end{eqnarray}  

By introducing the standard lengths $a_{\perp} = (\hbar/m\omega_{\perp})^{1/2}$ and $a_{z} = (\hbar/m\omega_{z})^{1/2}$, we can rescale the spatial coordinate, the energy, and the wave function as $\mathbf{r} = a_{\perp} \mathbf{r}_{1}$, $E = \hbar \omega_{\perp} E_{1}$ and $\psi(\mathbf{r}) = \sqrt{N/a^{3}_{\perp}} \psi_{1}(\mathbf{r}_{1})$, with $\omega_{x} = \omega_{y} = \omega_{\perp}$. Here the wave function $\psi_{1}$ is normalized to $1$. With help of the introduced asymmetry parameter $\lambda = \omega_{z}/\omega_{\perp}$ and the quantity $g = 4\pi a N/a_{\perp}$, the time-independent GP equation (\ref{TIGPE}) can be written as:
\begin{eqnarray}\label{TIGPE_r1}
 \Big[-\frac{1}{2} \nabla^{2}_{1} + \frac{(x^{2}_{1} + y^{2}_{1})}{2} + \frac{\lambda^{2} z^{2}_{1}}{2} + g |\psi(\mathbf{r}_{1})|^{2} \Big]\psi(\mathbf{r}_{1}) = \mu \psi(\mathbf{r}_{1}),
\end{eqnarray}

\section{Numerical prodecure}

In our calculation we use the Legendre-pseudospectral method \cite{Tsogbayar13}. In terms of this approach, function $\psi(r)$ can be expressed as:
\begin{eqnarray}\label{pseudo_eq_1}
 \psi(r) \approx \psi_{N_{r}}(r) = \sum^{N_{r}}_{i=0}\psi(r_{i}) g_{i}[x(r)],
\end{eqnarray}  
where $g_{i}[x(r)]$ is a cardinal function given with 
\begin{eqnarray}\label{pseudo_eq_2}
 g_{i}[x(r)] = -\frac{1}{N_{r}(N_{r}+1) P_{N_{r}} (x_{i})} \frac{(1-x^{2}) P^{'}_{N_{r}}(x)}{x - x_{i}}
\end{eqnarray}  
and $g_{i}(x_{j}) = \delta_{ij}$. Here $N_{r}$ is a number of grid point along $r(x) = a + (1+x)(b-a)/2, \, x\in [-1,1]$ and $r\in [a,b]$ with a length parameters $(a,b)$. Here the Legendre-Gauss-Lobatto grid pints $x_{i}$ are determined as the roots of the first derivative of the Legendre polynomial $P_{N}(x)$ with respect to $x$, $P'_{N_{r}}(x_{i}) = 0, \,\, i = 0, \ldots, N_{r}$. In the approach, the Laplace operator $\nabla^{2}$ can be approximated with a differentiation matrix $d_{ij}$~\cite{Tsogbayar13}. So, for the 3D calculation, we can approximate $\nabla^{2}_{x} + \nabla^{2}_{y} + \nabla^{2}_{z} \approx  I_{zz}\otimes(d^{2}_{xx}\otimes I_{yy}) + I_{zz}\otimes(I_{xx}\otimes d^{2}_{yy}) + (I_{xx}\otimes I_{yy})\otimes d^{2}_{zz}$. Here $I$ is the unit matrix, and $\otimes$ expresses the Kroneckor  (tensor) product~\cite{Tsogbayar13}. In our numerical calculation we use $a_{x}=a_{y}=a_{z}=a=-5$, $b_{x}=b_{y}=b_{z}=b=5$ in units of $a_{\perp}$, and $N_{x}=N_{y}=N_{z} = 32$.

\section{Results and discussion} 

As an example of atoms with repulsive interaction, we choose $^{87}\mbox{Rb}$, as in the experiment of Ref.~\cite{Anderson95}. In our calculation, all values of the physical parameters are taken from Ref.~\cite{Dalfovo96}: the $s$-wave triplet-spin scattering length, $a_{s} = 100a_{0}$ where $a_{0}$ is the Bohr radius; the asymmetry parameter of the experimental trap is $\lambda = \omega_{z}/\omega_{\perp} = \sqrt{8}$; the axial frequency $\omega_{z}/2\pi = 220\,Hz$; the corresponding characteristic length is $a_{\perp} = 1.222\times 10^{-12}\,cm$ and the ratio between the scattering and the oscillator lengths is $a/a_{\perp} = 4.33\times 10^{-3}$. In our calculation number of grid points is $N_{x} = N_{y} = N_{z} = 24$, and results are independent on this number. Table 1 shows the excess chemical potential and energy per particle for three values of $N=100, 1000$ and $1000$, and our calculated values are close to those in Ref.~\cite{Dalfovo96}, which had been obtained with a direct minimization approach combined with an imaginary time technique. Both quantities are expressed in units of $\hbar \omega_{\perp}$. 
\begin{table}[h]
\caption{ Results for the ground state of $^{87}\mbox{Rb}$ atoms in a trap with $\lambda = \sqrt{8}$. Chemical potential and energy are in unit of $\hbar \omega_{\perp}$. A number of grid point is $N_{x} = N_{y} = N_{z} = 24$. }
\begin{center}
{\scriptsize
\begin{tabular}{c@{\hspace{2mm}}c@{\hspace{2mm}}c@{\hspace{2mm}}c@{\hspace{2mm}}c@{\hspace{2mm}}c@{\hspace{2mm}}c@{\hspace{2mm}}c@{\hspace{2mm}}c@{\hspace{2mm}}c@{\hspace{2mm}}c@{\hspace{2mm}}c@{\hspace{2mm}}c@{\hspace{2mm}}c@{\hspace{2mm}} }
\hline\hline
& \multicolumn{2}{c}{$N = 100$} & \multicolumn{2}{c}{$N = 1000$} & \multicolumn{2}{c}{$N = 5000$}  \\
\cline{2-7}
 & $\mu$  & $E$ & $\mu$  & $E$ & $\mu$  & $E$ \\
\hline
This work & 2.88 & 2.67 & 4.77 & 3.84 & 8.15 & 6.13 \\
\cite{Dalfovo96}  & 2.88 & 2.66 & 4.77 & 3.84 & 8.14 & 6.12 \\
\hline\hline
\end{tabular} }
\end{center}
\end{table}
In Figure 1 we show plots of the wave function along the $x$ (panel a) and the $z$ axis (panel b) for four values of $N$. When $N$ increases the repulsion among the atoms tends to lower the central density, and expands the cloud of the atoms towards region where the trapping potential is higher. This results in increase of an energy per particle. 
\begin{figure}[h]
\centering
    \mbox{\subfigure[]{\includegraphics[width=0.4500\textwidth]{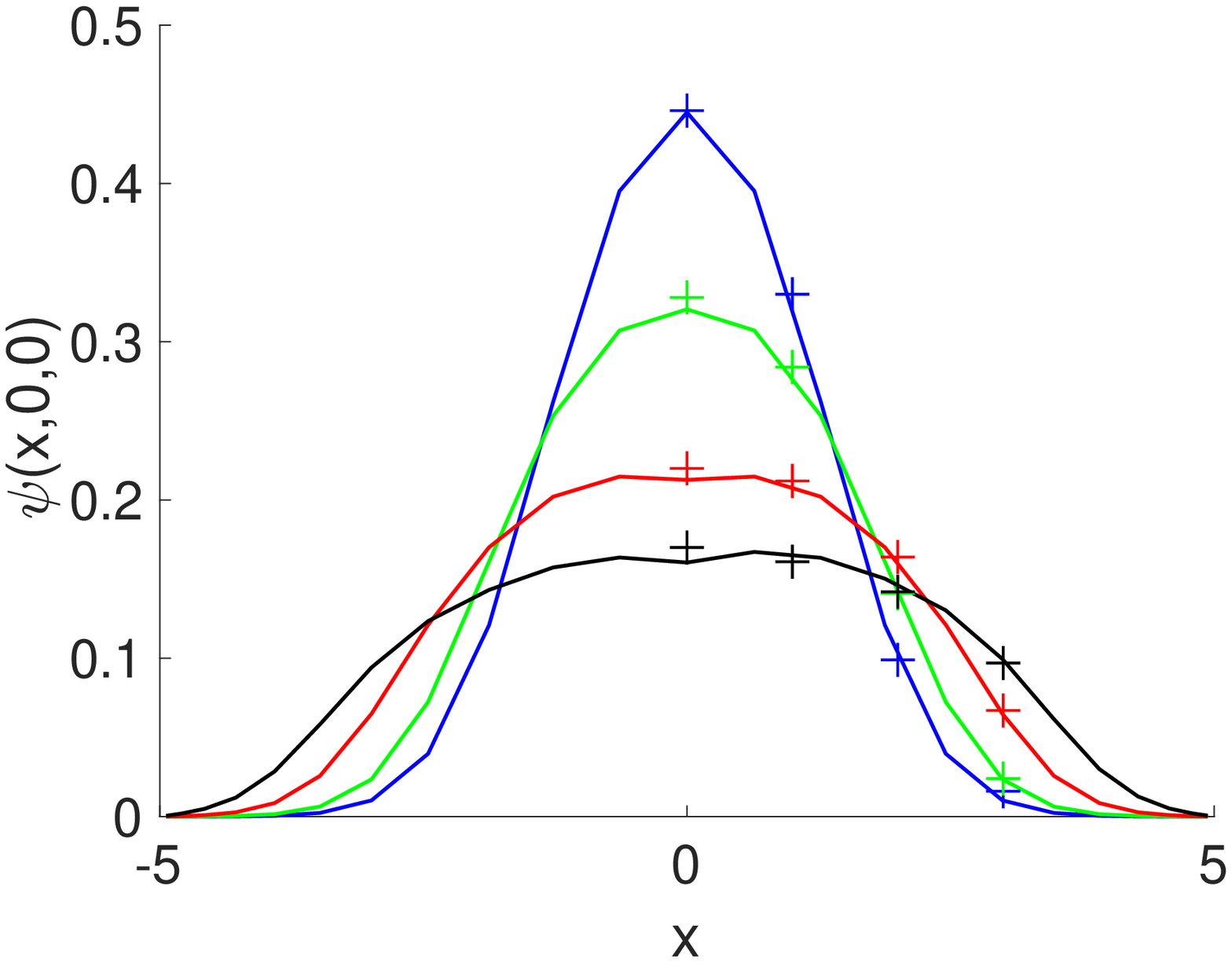}}
               \subfigure[]{\includegraphics[width=0.4500\textwidth]{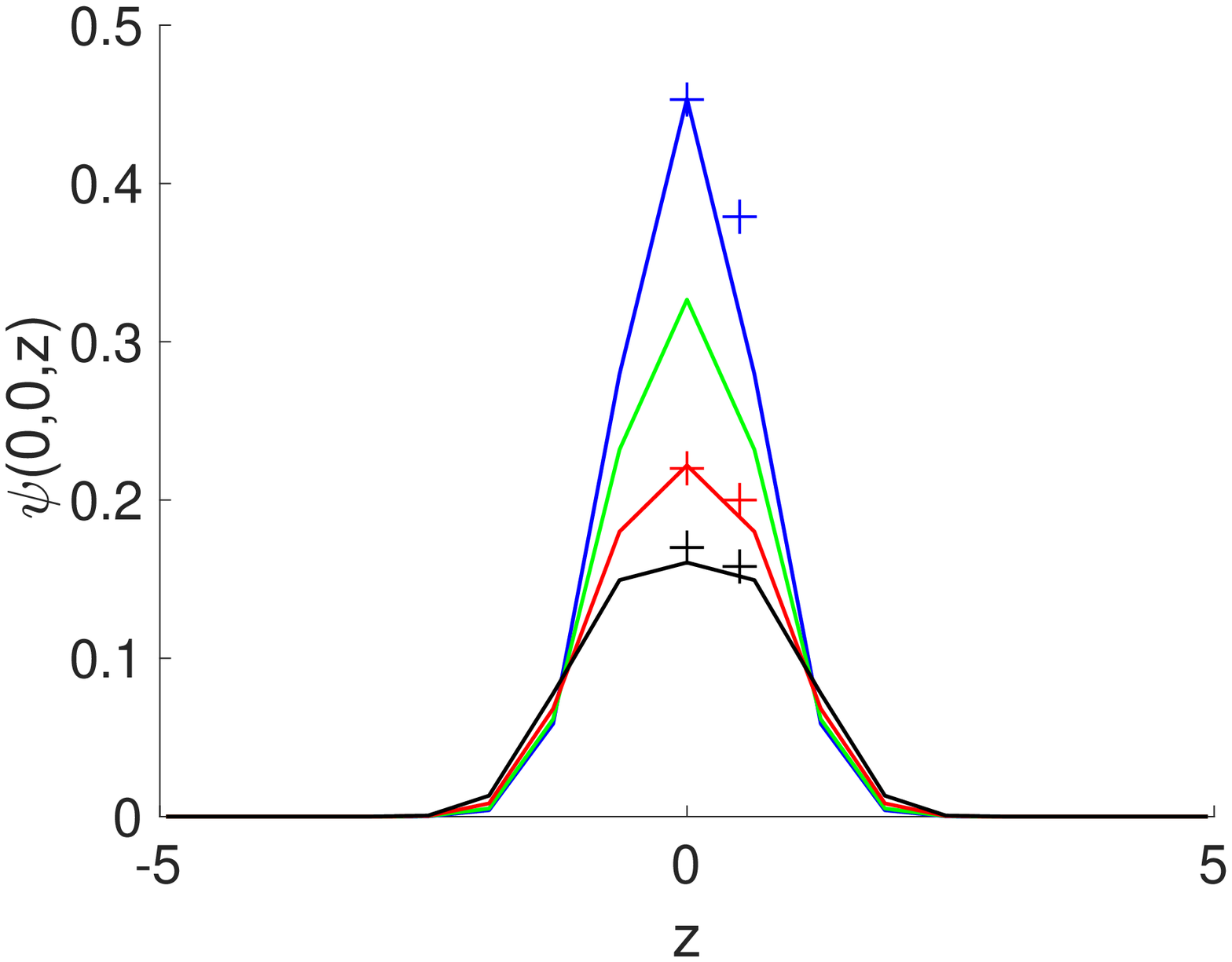}}
      }
\caption{ The ground-state wave function for $^{87}\mbox{Rb}$ along the $x$ axis (a) and along the $z$-axis (b). Distances are in units of $a_{\perp}$. Blue, green, red and black lines corresponds to $N=100, 500, 2000$ and $5000$, in descending order of central density. Cross are taken from Ref.~\cite{Dalfovo96}. } 
\end{figure}

As an example of atoms with an attractive interaction we choose $^{7}\mbox{Li}$, as in the experiment of Ref.~\cite{Bradley95}. In the calculation, we use $a_{s}=-27\,a_{0}$; $\omega_{z} = 2\pi\times 117$\,Hz; $a_{\perp} = 2.972\times 10^{-4}\,$cm; $|a|/a_{\perp} = 0.48\times 10^{-3}$; $\omega_{\perp} = 2\pi\times 163\,$Hz, and $\lambda = \omega_{z}/\omega_{\perp} = 0.72$. Table 2 shows numerical values of chemical potential and energy for the $^{7}\mbox{Li}$ ground state in unit of $\hbar \omega_{\perp}$ for three values of $N$. 
\begin{table}[h]
\caption{Same results as shown in Table 1, but for $^{7}\mbox{Li}$ atoms in a trap with $\lambda = 0.72$.}
\begin{center}
{\scriptsize
\begin{tabular}{c@{\hspace{2mm}}c@{\hspace{2mm}}c@{\hspace{2mm}}c@{\hspace{2mm}}c@{\hspace{2mm}}c@{\hspace{2mm}}c@{\hspace{2mm}}c@{\hspace{2mm}}c@{\hspace{2mm}}c@{\hspace{2mm}}c@{\hspace{2mm}}c@{\hspace{2mm}}c@{\hspace{2mm}}c@{\hspace{2mm}} }
\hline\hline
& \multicolumn{2}{c}{$N = 100$} & \multicolumn{2}{c}{$N = 500$} & \multicolumn{2}{c}{$N = 1000$}  \\
\cline{2-7}
 & $\mu$  & $E$ & $\mu$  & $E$ & $\mu$  & $E$ \\
\hline
This work & $1.33$ & $1.34$ & $1.17$ & $1.27$ & $0.86$ & $1.16$ \\
\hline\hline
\end{tabular} }
\end{center}
\end{table}
Figure 2a presents the ground state wave function for the $^{7}\mbox{Li}$ atom for three values of $N$. In this plot, the central density of cloud increases rapidly with $N$ since more attractive potential energy is added. Fig.~2b shows two-dimensional decsription of ground state wave function $\psi(x,y,0)$ for $N=1000$.  

\begin{figure}[h]
\centering
    \mbox{\subfigure[]{\includegraphics[width=0.4500\textwidth]{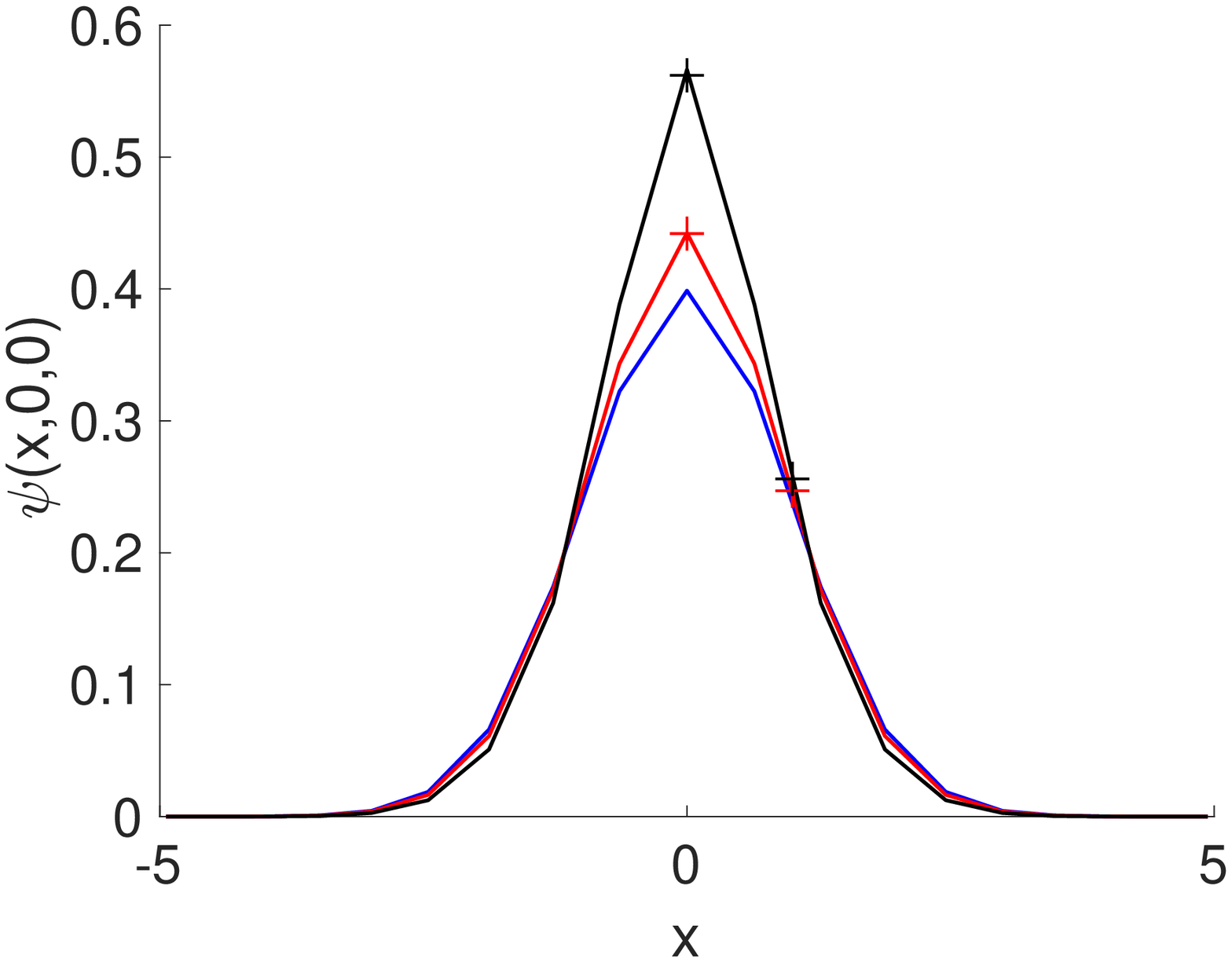}}
               \subfigure[]{\includegraphics[width=0.5000\textwidth]{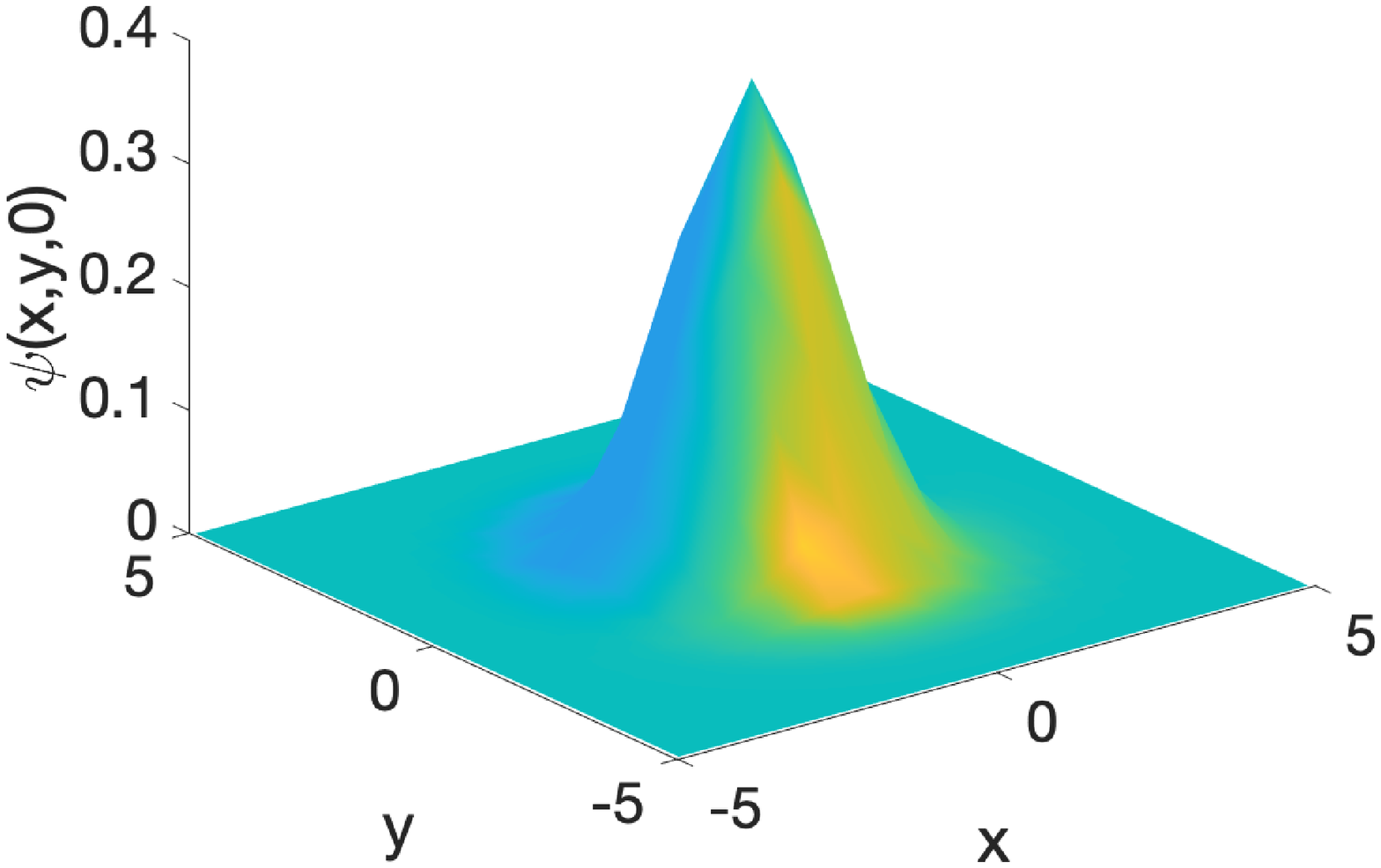}}
      }
\caption{ Same plots as shown in Figure 1, but for the $^{7}\mbox{Li}$ ground state along the $x$ axis (a), and blue, red and black lines correspond to $N=100, 500$ and $1000$, in ascending order of central density (a). Panel (b) shows an interpolated two-dimensional plot for $N=1000$. Crosses are taken from Ref.~\cite{Dalfovo96}.   } 
\end{figure}

\section{Conclusion} 

In this paper we have solved the time-independent Gross-Pitaevskii equation, non-linear eigenvalue problem, for a dilute gas of alkali atoms in an anisotropic traps using the pseudospectral method. The ground state wave function for the condensate with repulsive $(^{87}\mbox{Rb})$ and attractive ($^{7}\mbox{Li}$) behaviors at $T = 0$ has been obtained and natures of these wave function depending on number of particle have been discussed. Chemical potential and energy per particle for the $(^{87}\mbox{Rb})$ and ($^{7}\mbox{Li}$) condensates for different values of $N$ have been presented as well. 


\end{document}